\documentclass[prd,twocolumn,amsmath,amssymb]{revtex4}
\usepackage{graphicx}

\begin{document}

\title{\boldmath
Direct Measurement of the Branching Fraction for the Decay of
$D^+ \rightarrow \overline K^0 e^+\nu_e$ and Determination of
$\Gamma (D^0 \rightarrow K^-e^+\nu_e)/
\Gamma (D^+ \rightarrow \overline K^0 e^+\nu_e)$ }
\author{
\begin{small}
M.~Ablikim$^{1}$, J.~Z.~Bai$^{1}$, Y.~Ban$^{10}$,
J.~G.~Bian$^{1}$, X.~Cai$^{1}$, J.~F.~Chang$^{1}$,
H.~F.~Chen$^{15}$, H.~S.~Chen$^{1}$, H.~X.~Chen$^{1}$,
J.~C.~Chen$^{1}$
% ...
\end{small}
\\(BES Collaboration)\\
}
\vspace{0.2cm}
\affiliation{
\begin{minipage}{145mm}
$^{1}$ Institute of High Energy Physics, Beijing 100039, People's Republic
of China\\
$^{2}$ China Center for Advanced Science and Technology,
Beijing 100080, People's Republic of China\\
$^{3}$ Guangxi Normal University, Guilin 541004, People's Republic
of China\\
$^{4}$ Henan Normal University, Xinxiang 453002, People's Republic
of China\\
$^{5}$ Huazhong Normal University, Wuhan 430079, People's Republic
of China\\
$^{6}$ Hunan University, Changsha 410082, People's Republic of China\\
$^{7}$ Liaoning University, Shenyang 110036, People's Republic of China\\
$^{8}$ Nanjing Normal University, Nanjing 210097, People's Republic of
China\\
$^{9}$ Nankai University, Tianjin 300071, People's Republic of China\\
$^{10}$ Peking University, Beijing 100871, People's Republic of China\\
$^{11}$ Shandong University, Jinan 250100, People's Republic of China\\
$^{12}$ Shanghai Jiaotong University, Shanghai 200030, People's Republic
of China\\
$^{13}$ Sichuan University, Chengdu 610064, People's Republic of China\\
$^{14}$ Tsinghua University, Beijing 100084, People's Republic of China\\
$^{15}$ University of Science and Technology of China, Hefei 230026,
    People's Republic of China\\
$^{16}$ Wuhan University, Wuhan 430072, People's Republic of China\\
$^{17}$ Zhejiang University, Hangzhou 310028, People's Republic of China\\
\vspace{0.4cm}
\end{minipage}
}

\begin{abstract}    
The absolute branching fraction for the decay $D^+ \rightarrow
\overline K^0 e^+\nu _e$ is determined
using $5321\pm 149 \pm 160$ singly tagged $D^-$ event
sample from the data collected around 3.773 GeV
with the BES-II detector at the BEPC collider. In the system recoiling against
the singly tagged $D^-$ mesons, $34.4\pm 6.1$ events
for $D^+ \rightarrow \overline K^0 e ^+\nu _e$ are observed. 
Those yield the absolute
branching fraction to be
$BF(D^+ \rightarrow \overline K^0 e^+\nu_e)=(8.95 \pm 1.59\pm 0.67)\%$.
The ratio of the two partial widths
for the decays of
$D^0 \rightarrow K^-e^+\nu_e$ and $D^+ \rightarrow \overline K^0 e ^+\nu _e$
is determined to be
$ \Gamma (D^0 \rightarrow K^-e^+\nu_e)/
\Gamma (D^+ \rightarrow \overline K^0 e^+\nu_e) = 1.08\pm 0.22 \pm 0.07$.
\end{abstract}

\maketitle

\section{\bf Introduction} 
\begin{figure}[hbt]
\includegraphics[width=4.0cm,height=2.8cm]
{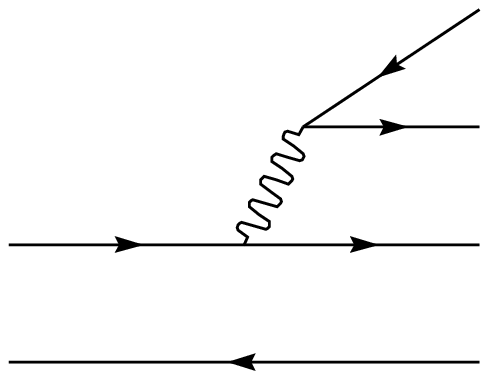}
\put(-115,27){\large $c$}
\put(0,27){\large $s$}
\put(-115,5){\large $\bar d$}
\put(-140,16.0){\large $D^+$}
\put(20,16.0){\large ${\overline K}^0$}
\put(0,5){\large $\bar d$}
\put(0,68){\large $e^+$}
\put(0,45){\large $\nu_e$}
\put(-75,36){\large $W^+$}
\put(-65,-10){(b)}
\caption{The decay diagrams for (a) $D^0 \rightarrow K^- e^+\nu_e$
and (b) $D^+ \rightarrow \overline K^0 e^+\nu_e$.
}
   \label{dcydgrm}
\end{figure}

Experimental study of the exclusive semileptonic decays of the charged and
neutral $D$ mesons can provide important information on the decay mechanisms.
Fig.~\ref{dcydgrm}(a) and Fig.~\ref{dcydgrm}(b) show the decay diagrams of the 
$D^0 \rightarrow K^-e^+\nu_e$ and the 
$D^+ \rightarrow \overline {K}^0 e^+\nu_e$, respectively. 
The isospin symmetry predicts that the partial widths of the two
decay processes should be equal. However, the ratio of the two partial
widths is 
$$\frac{\Gamma (D^0 \rightarrow K^-e^+\nu_e)}
{\Gamma (D^+ \rightarrow \overline{K}^0e^+\nu_e)} =1.34 \pm 0.20,$$
\noindent
which is determined based on the measured branching fractions
for the two decay processes and the lifetimes of the $D^0$ and $D^+$
quoted from PDG~\cite{pdg}. 
It deviates by $1.7 \sigma$ from the expected unit ratio.

To test the isospin symmetry in the exclusive semileptonic decays of
the charged and neutral $D$ mesons, 
we studied the two decay modes of 
$D^0 \rightarrow K^-e^+\nu_e$~\cite{smlp1_pbl}
and  $D^+ \rightarrow \overline{K}^0e^+\nu_e$
with the same data sample
of about 33 $\rm pb^{-1}$ collected 
at and around the center-of-mass energy of 3.773 GeV
with the BES-II detector at the BEPC collider.
In this Letter, we report the direct measurements of the branching fraction
for the decay of $D^+ \rightarrow \overline{K}^0e^+\nu_e$ (Throughout this
Letter, charged conjugation is implied) 
and the ratio of the two decay widths.

\begin{figure}[hbt]
\includegraphics[width=4.0cm,height=2.8cm]
{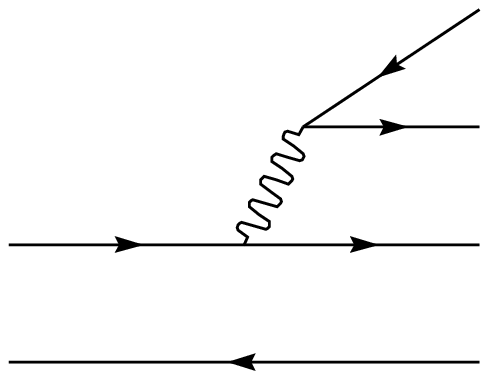}
\put(-115,27){\large $c$}
\put(0,27){\large $s$}
\put(-115,5){\large $\bar u$}
\put(-140,16.0){\large $D^0$}
\put(20,16.0){\large $K^-$}
\put(0,5){\large $\bar u$}
\put(0,68){\large $e^+$}
\put(0,45){\large $\nu_e$}
\put(-75,36){\large $W^+$}
\put(-65,-10){(a)}
\end{figure}

\section{\bf The BES-II DETECTOR}
The BES-II is a conventional cylindrical magnetic detector that is
described in detail in Ref.~\cite{bes}.  A 12-layer vertex chamber
(VC) surrounding the beryllium beam pipe provides input to the event
trigger, as well as coordinate information.  A forty-layer main drift
chamber (MDC) located just outside the VC yields precise measurements
of charged particle trajectories with a solid angle coverage of $85\%$
of $4\pi$; it also provides ionization energy loss ($dE/dx$)
measurements which are used for particle identification.  Momentum
resolution of $1.7\%\sqrt{1+p^2}$ ($p$ in GeV/c) and $dE/dx$
resolution of $8.5\%$ for Bhabha scattering electrons are obtained for
the data taken at $\sqrt{s}=3.773$ GeV. An array of 48 scintillation
counters surrounding the MDC measures the time of flight (TOF) of
charged particles with a resolution of about 180 ps for electrons.
Outside the TOF, a 12 radiation length, lead-gas barrel shower counter
(BSC), operating in limited streamer mode, measures the energies of
electrons and photons over $80\%$ of the total solid angle with an
energy resolution of $\sigma_E/E=0.22/\sqrt{E}$ ($E$ in GeV) and spatial
resolutions of
$\sigma_{\phi}=7.9$ mrad and $\sigma_Z=2.3$ cm for
electrons. A solenoidal magnet outside the BSC provides a 0.4 T
magnetic field in the central tracking region of the detector. Three
double-layer muon counters instrument the magnet flux return, and serve
to identify muons of momentum greater than 500 MeV/c. They cover
$68\%$ of the total solid angle.

\section{ DATA ANALYSIS}

Around the center of mass energy $3.773$ GeV, the
$\psi(3770)$ resonance is produced in electron-positron
($e^+e^-$) annihilation. The $\psi(3770)$
decays predominantly into $D\overline D$ pairs.
If a $D^-$ meson is fully reconstructed (This is called a
singly tagged $D^-$ meson),
the $D^+$ meson must exist
in the system recoiling against the singly tagged $D^-$ meson.
Using the singly tagged $D^-$ meson sample, the
decay of $D^+ \rightarrow \overline K^0 e^+\nu_e$
can be well selected in the recoiling system. 
Therefore, the absolute branching fraction for the decay of 
$D^+ \rightarrow \overline K^0 e^+\nu_e$ can be well measured.

\subsection{Events selection}

The $D^-$ meson is reconstructed in
non-leptonic decay modes of 
$K^+\pi^-\pi^-$, $K^0\pi^-$, $K^0K^-$, $K^+K^-\pi^-$,
$K^0\pi^-\pi^-\pi^+$, $K^0\pi^-\pi^0$,  $K^+\pi^-\pi^-\pi^0$,
$K^+\pi^+\pi^-\pi^-\pi^-$ and $\pi^+\pi^-\pi^-$.
Events which contain at least
three reconstructed charged tracks with good helix fits are selected.
In order to ensure
well-measured 3-momentum vectors and reliably charged particle
identification, the charged tracks used in the single tag analysis
are required to be within $|cos\theta|<$0.85, 
where $\theta$ is the polar angle.
All tracks, save those from $K^0_S$ decays, must originate
from the interaction region,
which require that the closest approach of a charged track
in the $xy$ plane is less than 2.0 cm and
the $z$ position of the charged track is less than 20.0 cm.
Pions and kaons are identified by means of
TOF and $dE/dx$ measurements. Pion identification requires a consistency
with the pion hypothesis at a confidence level ($CL_{\pi}$) greater than
$0.1\%$.
In order to reduce misidentification, a kaon candidate is
required to have a larger confidence level ($CL_{K}$) for a kaon hypothesis
than that for a pion hypothesis.
For electron identification,
the combined confidence level ($CL_{e}$), calculated for the $e$
hypothesis using
the $dE/dx$, TOF and BSC measurements, is required to be greater than
$0.1\%$, and the ratio
$CL_e/(CL_e + CL_{\pi} + CL_K)$ is required
to be greater than $0.8$.
The $\pi^0$ is reconstructed in the decay of
$\pi^0 \rightarrow \gamma\gamma$.
To select good photons from the decay
of $\pi^0$, the energy of a photon deposited in the BSC
is required to be greater than $0.07$ GeV~\cite{smlp1_pbl}, and the electromagnetic shower
is required to start in the first 5 readout layers. 
In order to reduce backgrounds, the angle between the
photon and the nearest charged track is required to be greater
than $22^{\circ}$~\cite{smlp1_pbl}
and the angle between the direction of the cluster development
and the direction of the photon emission to be less than
$37^{\circ}$~\cite{smlp1_pbl}.

For the single tag modes of $D^- \rightarrow K^+\pi^+\pi^-\pi^-\pi^-$ and
$D^- \rightarrow \pi^+\pi^-\pi^-$, backgrounds are further reduced by
requiring the difference between the measured energy
of the $D^-$ candidate and the beam energy
to be less than 70 and 60 MeV, respectively.
In addition, the cosine of the $D^-$ production angle relative to the beam
direction is required to be $|cos\theta_{D^-}|<0.8$.

\subsection{Singly tagged $D^-$ sample}
For each event, there may be several different charged track (or
charged and neutral track) combinations for each of
the nine single tag modes.
Each combination is subject to a one-constraint (1C) kinematic fit requiring
overall event energy conservation and that the unmeasured recoil system has
the same invariant mass as the track combinations.
Candidates with a fit probability $P(\chi^2)$ greater
than $0.1\%$ are retained.
If more than one combination satisfies $P(\chi^2)>0.1\%$,
the combination with the largest fit probability is retained.
For the single tag modes
with a neutral kaon and/or neutral pion,
one additional constraint kinematic fit
for the $K^0_S \rightarrow \pi^+\pi^-$ and/or
$\pi^0 \rightarrow \gamma\gamma$ hypothesis is performed, separately.

\begin{figure}[hbt]
\includegraphics[width=9.0cm,height=11.5cm]
{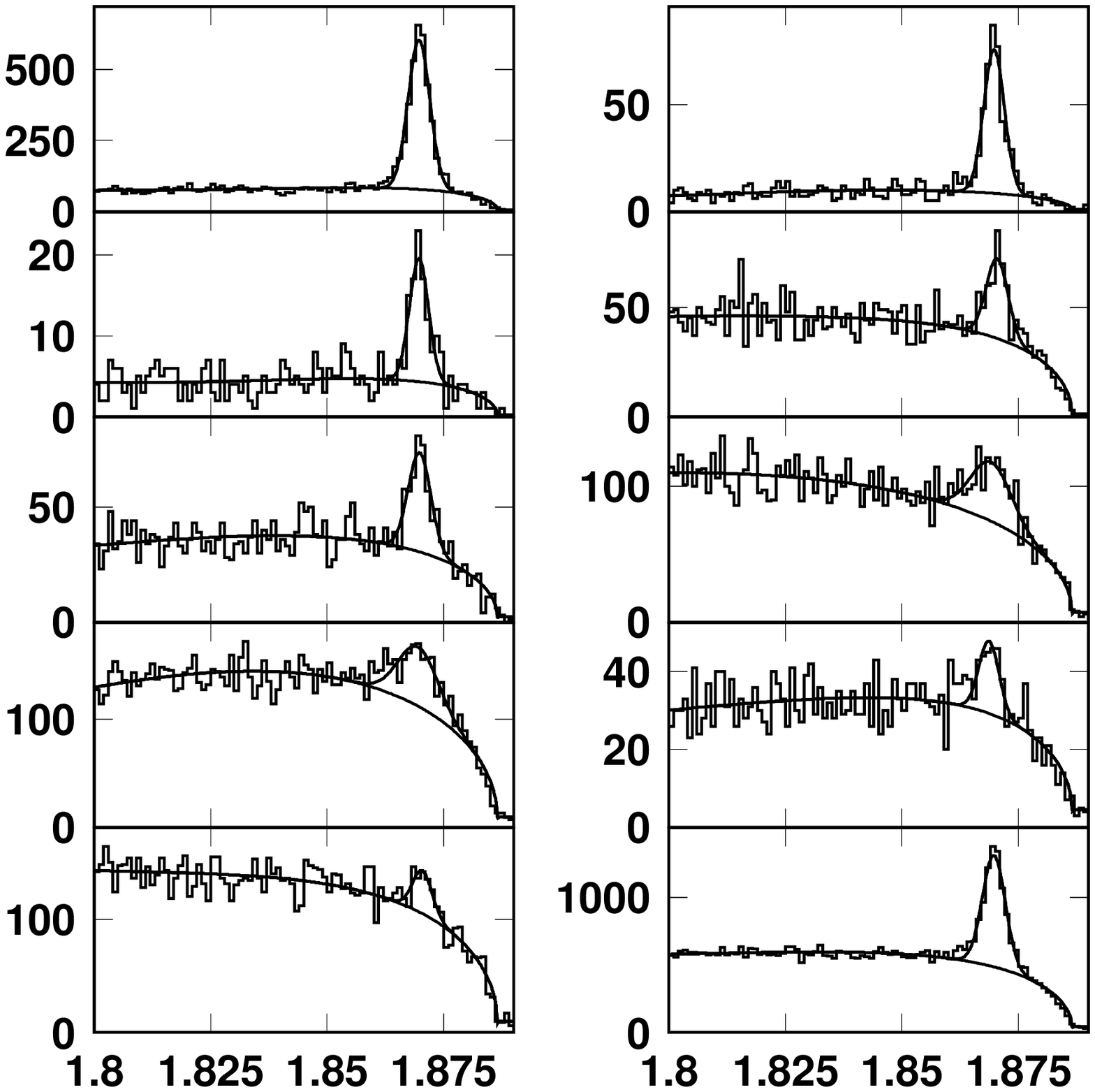}
\put(-165,5){Invariant Mass (GeV/$c^2$)}
\put(-260,130){\rotatebox{90}{Events/(0.001 GeV/$c^2$)}}
\put(-205,295){(a)}
\put(-90,295){(b)}
\put(-205,245){(c)}
\put(-90,245){(d)}
\put(-205,155){(e)}
\put(-90,155){(f)}
\put(-205,105){(g)}
\put(-90,105){(h)}
\put(-205,50){(i)}
\put(-90,50){(j)}
\caption{
The distributions of the fitted masses of
(a) $K^+\pi^-\pi^-$,
(b) $K^0\pi^-$,
(c) $K^0K^-$,
(d) $K^+K^-\pi^-$,
(e) $K^0\pi^-\pi^-\pi^+$,
(f) $K^0\pi^-\pi^0$,
(g) $K^+\pi^-\pi^-\pi^0$,
(h) $K^+\pi^+\pi^-\pi^-\pi^-$,
and
(i) $\pi^-\pi^-\pi^+$ combinations;
(j) is the fitted masses of the $mKn\pi$
combinations for the 9 modes combined together.
}
\label{sgltg}
\end{figure}

The resulting distributions in the fitted invariant masses
of $mKn\pi$ 
($m=0~ {\rm or}~1~{\rm or}~2$ and $n=1~{\rm or}~2~{\rm or}~3~{\rm or}~4$) 
combinations,
which are calculated using the fitted momentum vectors from the kinematic
fit, are shown in Fig.~\ref{sgltg}.
The signals for the singly tagged $D^-$ mesons are clearly observed
in the fitted mass spectra.
A maximum likelihood fit
to the mass spectrum with a Gaussian function for the $D^-$
signal and a special
background 
function~\footnote{A Gaussian function was assumed for the signal. The
background shape was
$$(1.0+p_1 y + p_2 y^2)
N\sqrt{1-(\frac{x}{E_b})^2}~x~e^{-f(1-\frac{x}{E_b})^2}+c,$$
where $N\sqrt{1-(\frac{x}{E_b})^2}~x~e^{-f(1-\frac{x}{E_b})^2}$
is the ARGUS background shape,
$x$ is the fitted mass, $E_b$ is the beam energy,
$y=(E_b-x)/(E_b -1.8)$,
$N$, $f$, $p_1$, $p_2$ and c are the fit parameters.
The parameter c accounts for the varying of the beam energy.
The ARGUS background shape was used by the ARGUS experiment to parameterize
the
background for fitting $B$ mass peaks. For details, 
see~\cite{mnfit}.
} 
to describe backgrounds
yields the number of the singly tagged $D^-$ events
for each of the nine modes and the total number of $5321\pm 149 \pm 160$
reconstructed $D^-$ mesons, where the
first error is statistical and the second systematic obtained
by varying the parameterization of the background.
The curves of Fig.~\ref{sgltg} give the best fits to the invariant mass
spectra. In the fits to the mass spectra, the standard deviations of the
Gaussian signal functions for the Fig.~\ref{sgltg}(g) and
Fig.~\ref{sgltg}(h)
are fixed at 4.27 MeV and 2.16 MeV, respectively.
These standard deviations are obtained from Monte Carlo
sample. All other parameters are left free in the fit.

\subsection{Candidates of $D^+ \rightarrow \overline K^0 e^+\nu_e$}
Candidate events of the decay $D^+ \rightarrow \overline K^0 e^+\nu_e$ 
are selected from the surviving tracks in the system recoiling against the
tagged $D^-$. To select the $D^+ \rightarrow \overline K^0 e^+\nu_e$,
it is required that there are three charged tracks,
one of which is identified as an electron with charge opposite to the charge
of the tagged $D^-$ and
the other two as $\pi^+$ and $\pi^-$. 
The difference between the invariant masses of the $\pi^+\pi^-$ combinations
and the mass of $K^0_S$ should be less than $20$ MeV/$c^2$.
The  neutrino is undetected, therefore the kinematic
quantity  $U_{miss}\equiv E_{miss}-p_{miss}$ is used
to obtain the information about the missing neutrino, where $E_{miss}$ and
$p_{miss}$ are the total energy
and the total momentum of all missing particles respectively,
which are carried by the undetected particles.
The backgrounds from the decays such as 
$D^+ \rightarrow \overline K^0 \pi^+ \pi^0$ and
$D^+ \rightarrow \overline K^0 \pi^0 e^+\nu_e$
are suppressed by rejecting the events with extra isolated
photons which are not used in the reconstruction 
of the singly tagged $D^-$ meson.
The isolated photon should have its energy greater
than 0.1 GeV~\cite{smlp1_pbl} and
satisfy photon selection criteria as mentioned earlier.

\begin{figure}[hbt]
\includegraphics[width=9.0cm,height=7.0cm]
{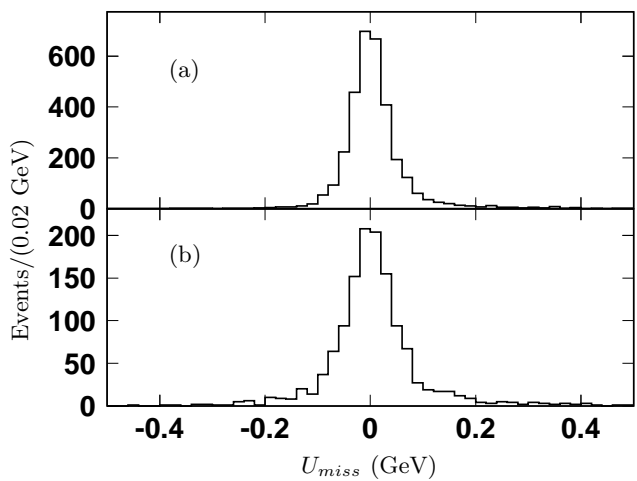}
\put(-140,10){$U_{miss}$ (GeV)}
\put(-250,60){\rotatebox{90}{Events/(0.02 GeV)}}
\put(-190,160){(a)}
\put(-190,90){(b)}
\caption{The distributions of $U_{miss}$ calculated for the Monte Carlo events of
(a) $D^+ \rightarrow {\overline K}^0 e^+ \nu$ versus $D^- \rightarrow K^+\pi^-\pi^-$
and (b) $D^+ \rightarrow {\overline K}^0 e^+ \nu$ versus
$D^- \rightarrow K^+\pi^-\pi^-\pi^0$
.
}
\label{umiss_mc}
\end{figure}

\begin{figure}[hbt]
\includegraphics[width=9.0cm,height=6.0cm]
{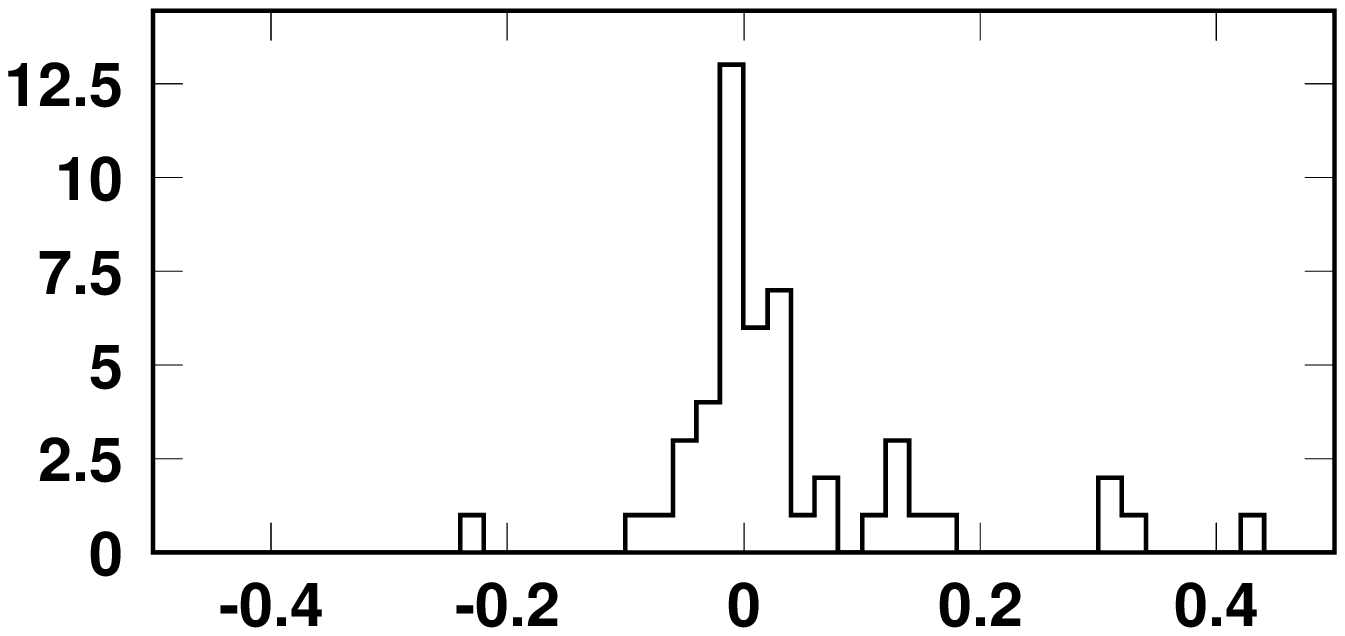}
\put(-140,10){$U_{miss}$ (GeV)}
\put(-250,60){\rotatebox{90}{Events/(0.02 GeV)}}
\caption{The distribution of $U_{miss}$ calculated for the selected events of
$D^+ \rightarrow {\overline K}^0 e^+ \nu$ versus $D^-$
tags ($mKn\pi$ combinations).
}
\label{umiss_data}
\end{figure}

\begin{figure}[hbt]
\includegraphics[width=9.0cm,height=6.0cm]
{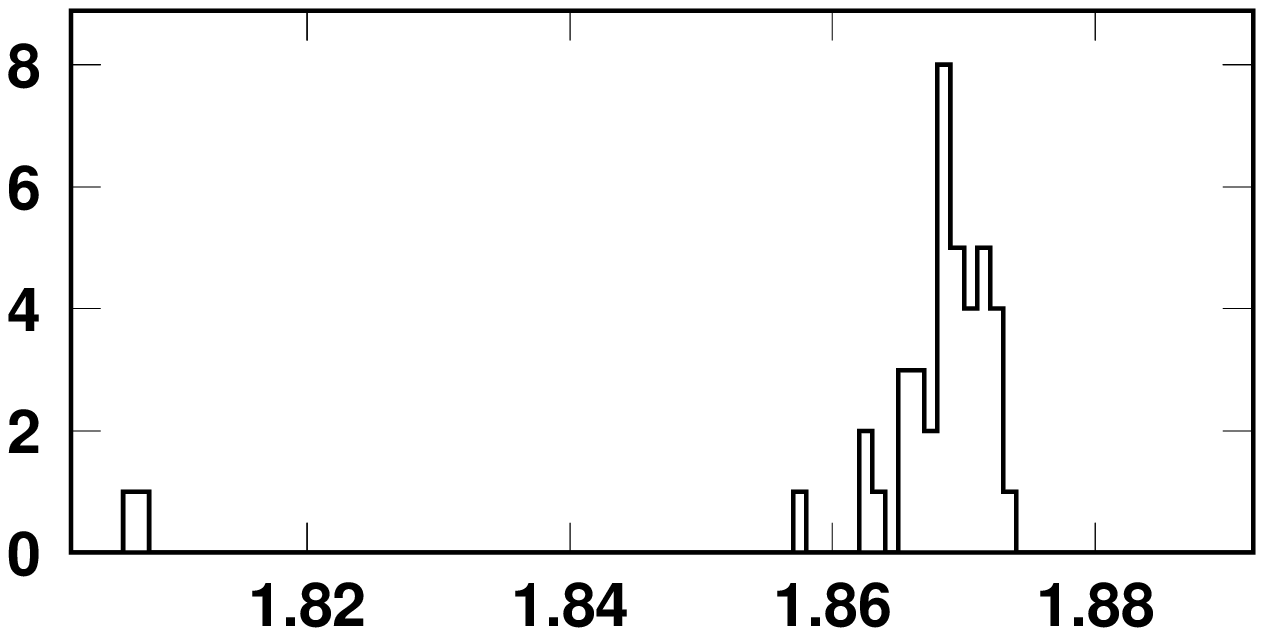}
\put(-155,10){Invariant Mass (GeV/$c^2$)}
\put(-250,60){\rotatebox{90}{Events/(0.001 GeV/$c^2$)}}
\caption{The distribution of the fitted invariant masses of $mKn\pi$
combinations for the events in which the 
$D^+ \rightarrow {\overline K}^0e^+\nu_e$ candidate events 
are observed in the system recoiling against the $mKn\pi$ combinations; where a clear
signal for the singly tagged $D^-$ is observed. 
}
\label{dbtag}
\end{figure}

Fig.~\ref{umiss_mc} shows the $U_{miss}$ distributions for
the Monte Carlo $D^+ \rightarrow \overline K^0 e^+\nu_e$ 
events. 
The quantity $U_{miss}$ is close to zero as expected.
Fig.~\ref{umiss_data} shows the $U_{miss}$ distribution for the events from
the data, which satisfy the selection criteria. 
From the distribution, most of the events 
can be identified as the candidates of 
$D^+ \rightarrow \overline K^0 e^+\nu_e$ versus 
the tagged $D^-$ mesons. Those can
be further confirmed as follows. 
If we select the events which
satisfy the requirement $|U_{miss,i}|<3 \sigma_{U_{miss,i}}$ 
for the single tag mode($i$),
where $\sigma_{U_{miss,i}}$ is the standard deviation of the $U_{miss,i}$
distribution obtained from
Monte Carlo simulation for the event of
$D^+ \rightarrow \overline K^0 e^+\nu_e$ 
versus the single tag mode($i$) ($i=1$ is
for $K^+\pi^-\pi^-$; $i=2$ is for $K^0\pi^-$... 
and $i=9$ is for $\pi^+\pi^-\pi^-$ modes),
and plot the fitted masses of the $mKn\pi$ combinations, 
we observe a clear signal of $D^+ \rightarrow \overline K^0 e^+\nu _e$ 
versus the $D^-$ tags as shown in Fig.~\ref{dbtag}.
In Fig.~\ref{dbtag}, there are 37 events
in the $\pm 3\sigma_{mass,i}$ signal regions,
while there are 4 events in the
outside of the signal regions;
where the $\sigma_{mass,i}$ is the standard deviation of the fitted mass
distribution for the single tag mode(i).
By assuming that the background distribution is flat,
$0.8 \pm 0.4$ background events are estimated in the signal region.
There may also be the $\pi^+\pi^-$ combinatorial background. 
By selecting the events in which the invariant masses of the $\pi^+\pi^-$ 
combinations in the recoil side of the tags
are outside of the $K^0_S$ mass window, 
we estimate that there are
$0.3 \pm 0.2$ background events in the candidate events. 
After subtracting these numbers of background events,
$35.9 \pm 6.1$ candidate events are retained.

\begin{figure}[hbt]
\includegraphics[width=9.0cm,height=6.0cm]
{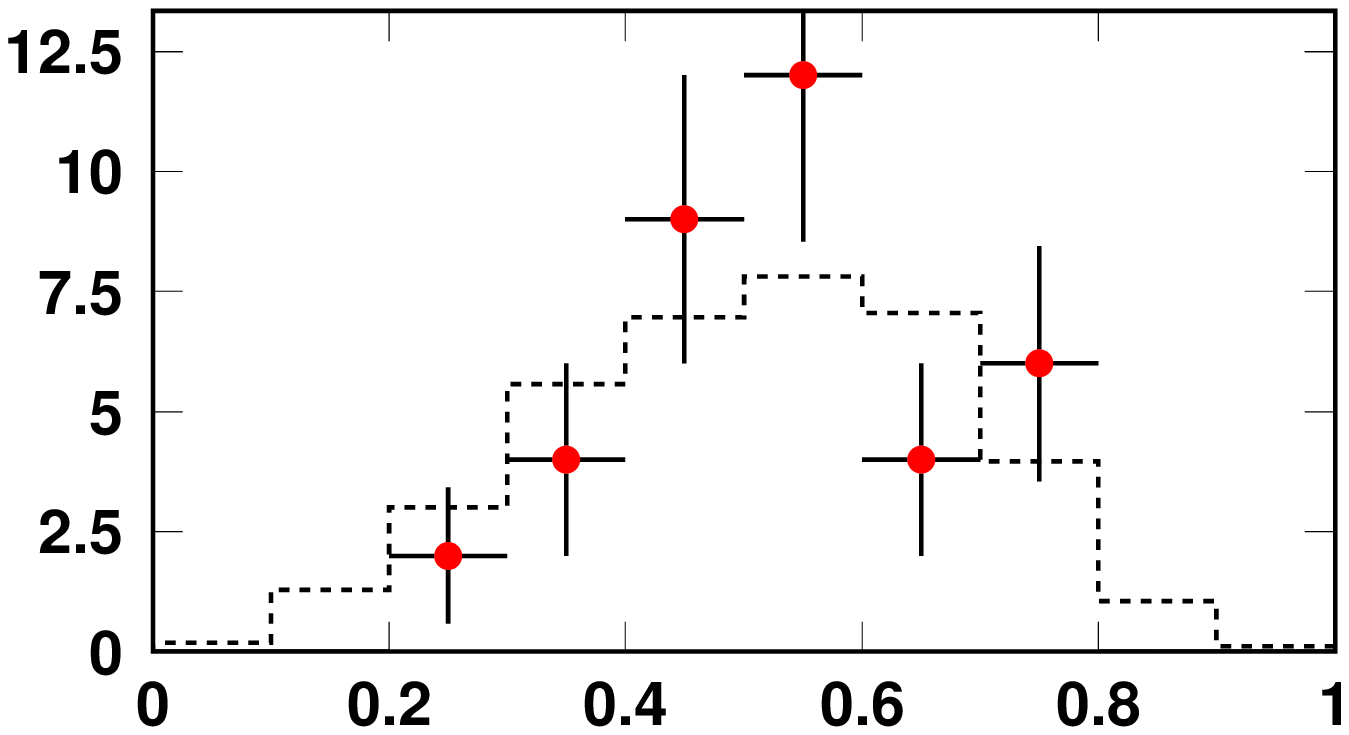}
\put(-160,5){Momentum of electron (GeV/$c$)}
\put(-250,60){\rotatebox{90}{Events/(0.1 GeV/$c$)}}
\caption{The distribution of the momentum of the electrons from the
selected candidate events of $D^+ \rightarrow {\overline K}^0e^+\nu_e$, 
where the error bars are for the events from the
data and the histogram is for the events of 
$D^+ \rightarrow {\overline K}^0e^+\nu_e$ from the Monte Carlo sample.
}
\label{momentum_e}
\end{figure}

The distribution of the momentum of the electrons 
from the selected candidate events of
$D^+ \rightarrow {\overline K}^0e^+\nu_e$ is shown in
Fig.~\ref{momentum_e}, where the error bars 
are for the events from the data and the histogram
is for the events of $D^+ \rightarrow {\overline K}^0e^+\nu_e$
from Monte Carlo sample. The measured electron momentum is 
in good agreement
with the electron momentum from the Monte Carlo events of $D^+ \rightarrow
{\overline K}^0e^+\nu_e$.

\subsection{Background Subtraction}

There are still some background contaminations in the observed
candidate events due to other semileptonic or
hadronic decays. These background events must be subtracted
from the candidate events.
The numbers of background events are estimated by analyzing the
Monte Carlo sample which is 13 times larger than the data.
The Monte Carlo events are generated as
$e^+e^- \rightarrow D \overline D$ and the $D$ and $\overline D$
mesons are set to decay to all possible final states
according to the decay modes and branching fractions
quoted from PDG~\cite{pdg} excluding the decay mode under study.
The number of events satisfying the selection criteria
is then renormalized
to the corresponding data set.
Totally $1.5 \pm 0.4$ background
events are obtained for
$D^+ \rightarrow \overline K^0 e^+\nu_e$.
After subtracting the number of background events,
$34.4\pm 6.1$ signal events
for $D^+ \rightarrow \overline K^0 e^+\nu _e$
decay are retained.

\section{Results}

\subsection{Monte Carlo Efficiency}
The efficiency for reconstruction of the semileptonic decay
events of $D^+ \rightarrow \overline K^0 e^+\nu _e$
is estimated
by Monte Carlo simulation. A detailed Monte Carlo study gives
the efficiency to be
$\epsilon_{\overline K^0 e^+\nu _e}=(7.22 \pm 0.06)\%$,
where the error is statistical.

\subsection{Branching Fraction}

The measured branching fraction is obtained by dividing
the observed number of the semileptonic decay events
$N(D^+ \rightarrow \overline K^0 e^+ \nu_e)$
by the number of the singly tagged $D^-$
mesons $N_{D^-_{tag}}$ and the reconstruction efficiency
$\epsilon_{\overline K^0 e^+\nu_e}$,
\begin{equation}
Br(D^+ \rightarrow \overline K^0 e^+\nu _e)=
\frac{ N(D^+ \rightarrow \overline K^0 e^+ \nu_e) }
{ \epsilon_{\overline K^0 e^+ \nu_e} \times N_{D^-_{tag}}}.
\end{equation}
Inserting these numbers
into the equation (1),
the branching fraction
for $D^+\rightarrow \overline K^0 e^+\nu_e$
decay is obtained to be
$$BF(D^+ \rightarrow \overline K^0 e^+\nu_e)=(8.95\pm 1.59 \pm 0.67)\%,$$
where the first error is statistical and the second systematic.
The systematic uncertainty in the measured branching
fraction arises from the
particle identification $(1.5\%)$, tracking efficiency
($2.0\%$ per track), photon reconstruction ($2.0\%$),
$U_{miss}$ selection ($0.6\%$),
the number of the singly tagged $D^-$ mesons
($3.0\%$), background subtraction
($1.6\%$), Monte Carlo statistics ($0.7\%$) and
$K^0_S$ selection ($1.1\%$). 
These uncertainties are added in quadrature to obtain the total systematic
error, which is $7.5\%$.

Table I gives the comparison of our measured value of the branching fraction
for the decay of $D^+ \rightarrow {\overline K}^0 e^+ \nu_e$ with that
measured by MARK-III~\cite{mark3}.
Our measured branching fraction is consistent within the error
with that measured by MARK-III. 
\begin{table}
\caption{Comparison of our measured branching fraction 
for the decay of $D^+ \rightarrow {\overline K}^0 e^+ \nu_e$
with that measured by MARK-III.}
\begin{center}
\begin{tabular}{cc}
\hline \hline
Branching fraction~~ [$\%$] & Branching fraction~~ [$\%$] \\
 (This experiment)           & (MARK-III)  \\
\hline
 $8.95 \pm 1.59 \pm 0.67$  &
   $6.0^{+2.2}_{-1.3} \pm 0.7$  \\
\hline
\hline
\end{tabular}
\end{center}
\end{table}

\subsection{The ratio of $\Gamma(D^0 \rightarrow K^-e^+\nu_e)/\Gamma(D^+
\rightarrow \overline K^0 e^+ \nu_e)$}

With the same data sample, BES-II measured the absolute
branching fraction for $D^0 \to K^-e^+\nu _e$ decay to be
$BF(D^0 \rightarrow K^-e^+\nu_e)=
(3.82 \pm 0.40\pm 0.27)\%$~\cite{smlp1_pbl}. 
Using the measured branching fractions
for the decays of $D^0 \rightarrow K^-e^+\nu_e$ and 
$D^+ \rightarrow \overline K^0 e^+ \nu_e$ and the lifetimes of the
$D^0$ and $D^+$ quoted from the PDG~\cite{pdg}, the ratio of the decay
widths is obtained to be
$$\frac{\Gamma(D^0 \rightarrow K^-e^+\nu_e)}
{\Gamma(D^+ \rightarrow \overline K^0 e^+\nu_e)} =
1.08 \pm 0.22 \pm 0.07, $$
where the first error is statistical and the second systematic
which arises from some uncanceled systematic uncertainty ($6.8\%$)
in the measured ratio of the branching fractions 
for the two decay modes and the uncertainty ($0.8\%$) in the measured 
ratio of the $D^0$ and $D^+$ lifetimes~\cite{pdg}.

\section{Summary}

In summary, by analyzing the data sample of about $33$ $\rm {pb^{-1}}$ 
collected at and around 3.773 GeV
with the BES-II detector at the BEPC collider, 
the branching fraction for the decay
of $D^+ \rightarrow \overline K^0 e^+\nu_e$
has been measured. From a total of $5321\pm 149 \pm 160$
singly tagged $D^-$ event sample,
$34.4\pm 6.1$ $D^+ \rightarrow \overline K^0 e^+\nu_e$
signal events are observed in the system recoiling against the 
singly tagged $D^-$ mesons.
Those yield the decay branching fraction to be
$BF(D^+ \rightarrow \overline K^0 e^+\nu_e)=(8.95\pm 1.59\pm 0.67)\%$.
Using the values of the measured branching fractions for the decays
of  $D^0 \rightarrow K^- e^+\nu _e$ and
$D^+ \rightarrow \overline K^0 e^+\nu _e$,
the ratio of the two partial decay widths is determined to be
$$\frac{\Gamma(D^0 \rightarrow K^-e^+\nu_e)}
{\Gamma(D^+ \rightarrow \overline K^0 e^+\nu_e)} =
1.08 \pm 0.22 \pm 0.07,$$
which is consistent within the errors with
the theoretical prediction of the spectator model and 
supports that isospin conservation holds in the exclusive semileptonic decays
of the $D^+ \rightarrow \overline K^0 e^+\nu_e$ 
and the $D^0  \rightarrow K^- e^+\nu_e$. 

\vspace{5mm}

\begin{center}
{\small {\bf ACKNOWLEDGEMENTS}}
\end{center}
\par
\vspace{0.4cm}

   The BES Collaboration thanks the staff of BEPC for their hard efforts.
This work is supported in part by the National Natural Science Foundation
of China under contracts
Nos. 19991480,10225524,10225525, the Chinese Academy
of Sciences under contract No. KJ 95T-03, the 100 Talents Program of CAS
under Contract Nos. U-11, U-24, U-25, and the Knowledge Innovation Project
of CAS under Contract Nos. U-602, U-34(IHEP); by the
National Natural Science
Foundation of China under Contract No.10175060(USTC),and
No.10225522(Tsinghua University).


\begin{thebibliography}{9}
\bibitem{pdg} Particle Data Group, S. Eidelman {\it et al.}, 
                   Phys. Lett. B 592 (2004) 1.
\bibitem{smlp1_pbl} BES Collaboration, M. Ablikim {\it et al.}, 
                   Phys. Lett. B 597 (2004) 39, arXiv:hep-ex/0406028.
\bibitem{bes} BES Collaboration, J.Z. Bai {\it et al.}, 
                   Nucl. Instr. Meth. A 458 (2001) 627.
\bibitem{mnfit} Ian C. Brock, Mn-Fit, a fitting and plotting package using MINUIT, 
                Version 4.07, December 22, 2000.
\bibitem{mark3}MARK-III Collaboration, Z. Bai {\it et al.}, Phys.
Rev. Lett. 66 (1991) 1011.

\end{thebibliography}
\end{document}